\documentclass[aps,prb,twocolumn,showpacs]{revtex4}
\usepackage{graphicx}

\begin{document}



\title{Electronic structures of UTSn (T=Ni, Pd) using
	photoemission spectroscopy} 

\author {J.-S. Kang$^1$, S. C. Wi$^1$, J. H. Kim$^1$, K. A. McEwen$^2$,
	C. G. Olson$^3$, J. H. Shim$^4$, and B. I. Min$^4$}

\affiliation{$^1$Department of Physics, The Catholic University of Korea,
	Puchon 420-743, Korea}

\affiliation{$^2$Department of Physics and Astronomy, 
	University College London, WCIE 6BT, UK}

\affiliation{$^3$Ames Laboratory, Iowa State University,
	Ames, Iowa 50011, U.S.A.}

\affiliation{$^4$Department of Physics, Pohang University of Science and 
	Technology, Pohang 790-784, Korea}

\date{\today}

\begin{abstract}

The electronic structures of the localized $5f$ systems UTSn (T=Ni, Pd)
have been investigated using photoemission spectroscopy (PES).
The extracted U $5f$ PES spectra of UTSn (T=Ni, Pd) exhibit
a broad peak centered at $\sim 0.3$ eV below $\rm E_F$
with rather small spectral weight near $\rm E_F$ (N$_f$($\rm E_F$)).
The small N$_f$($\rm E_F$) in UTSn is found to be correlated with
the T $d$ PES spectra that have a very low density of states 
(DOS) near $\rm E_F$.
The high-resolution PES spectra for UTSn provide the V-shaped reduced 
metallic DOS near $\rm E_F$ but do not reveal any appreciable changes 
in their electronic structures across the magnetic phase transitions.
A possible origin for the reduced N$_f$($\rm E_F$) in UTSn is ascribed 
to the hybridization to the very low T $d$ DOS at $\rm E_F$.
Comparison of the measured PES spectra to the LSDA+$U$ band
structure calculation reveals a reasonably good agreement for UPdSn,
but not so for UNiSn.

\end{abstract} 

   
\maketitle


\section{Introduction}
\label{sec:intro}

Uranium intermetallic compounds often exhibit interesting magnetic
behavior that is neither very localized nor very itinerant.
UNiSn and UPdSn are considered to be well localized with small linear
specific-heat coefficient,
$\gamma \approx 18-28$ mJ/mol K$^2$ \cite{Palstra87} and
$\gamma \approx 5$ mJ/mol K$^2$ \cite{Boer92}, respectively,
and with the large ordered magnetic moments of 
$\sim 1.55~\mu_B$ \cite{Kawa89}
and $\sim 2~\mu_B$ \cite{Robinson}, respectively,
which are significantly larger than in other U intermetallic systems.
Both UNiSn and UPdSn exhibit interesting phase transitions with
the antiferromagnetic (AF) ground states.
UNiSn displays an AF order below the Neel temperature $\rm T_N
\simeq 43$ K,  a structural transition from the cubic 
symmetry to the tetragonal symmetry at $\rm T_N$ \cite{Akazawa96},
and a semiconductor-to-metal (S-M) transition around 
$\rm T_{MI} \sim 55$ K \cite{Fujii89,Diel93}.
This multiple phase transition seems to be anomalous because it is 
an inverse metal-insulator transition with a gap-opening 
above $\rm T_N$ and the structural, S-M, and AFM transitions occur 
concomitantly.
Hexagonal UPdSn also displays two AF transitions with concomitant
lattice distortions.
UPdSn undergoes an AF transition below $\rm T_N \simeq 37$ K with 
the orthorhombic magnetic symmetry (phase I),
and undergoes another AFM transition below $\simeq 25$ K
with the monoclinic magnetic symmetry (phase II) \cite{Nakotte98}.
Both AF structures in UPdSn are reported to be noncollinear.
The resistivity of UPdSn shows a metallic behavior in the whole 
temperature range but with a feature of rapid drop 
below $\rm T_N \simeq 37$ K \cite{Palstra87,Boer92}.

The underlying mechanism of the peculiar multiple phase transitions
in UNiSn and UPdSn has been investigated extensively
\cite{Aoki93,Akazawa98,McEwen00}.
A quadrupolar ordered phase based on the crystalline electric field
(CEF) level scheme for the localized $5f^2$ (U$^{4+}$) configuration
\cite{Aoki93} has been proposed for the phase transitions in UTSn
\cite{Akazawa98}, which seems to be consistent with the neutron inelastic
scattering data \cite{McEwen00}.
In contrast, the localized $5f^3$ (U$^{3+}$) configuration was
proposed based on the neutron diffraction data for UPdSn \cite{Johnson93}.
Some electronic structure calculation for UPdSn supports 
the localized U $5f^2$ configuration \cite{Trygg94},
whereas other calculations argue the itinerant character of U $5f$ 
electrons in UNiSn \cite{Oppen96} and UPdSn \cite{Sand97}.
Neither the theoretically predicted electronic structure of UTSn 
nor the $5f^2$ configuration of the localized U$^{4+}$ ion has 
been verified by photoemission spectroscopy (PES) experiment.
An early resonant photoemission spectroscopy (RPES) study on UTSn 
(T=Ni, Pd, Pt) with a rather poor instrumental resolution 
\cite{Hochst86} did not address the origin of the phase 
transitions in UTSn.
A recent PES study on UNiSn by some of the present authors 
\cite{Kang01} has found the importance of the on-site Coulomb 
interaction $U$ between U $5f$ electrons.
Therefore the nature of $5f$ electrons in UTSn is still controversial. 

In order to explore the role of the electronic structures
in the phase transitions of UTSn,
we have performed the RPES measurements of UTSn (T=Ni, Pd) near 
the U $5d \rightarrow 5f$ absorption edge and determined the partial 
spectral weight (PSW) distributions of both the U $5f$ and Ni/Pd $d$ 
electrons.
We have then compared the experimental data to the electronic structure 
calculation performed in the LSDA+$U$ method 
(LSDA: local spin-density functional approximation) \cite{Anisimov97}.

\section{Experimental and Calculational Details}
\label{sec:exp}

UNiSn and UPdSn polycrystalline samples are made by arc melting
constituent elements of high purity \cite{Kang01}.
Our magnetization measurements taken on the sample after one month 
annealing showed clear antiferromagnetic transitions
in agreement with previous results \cite{Fujii89,Nakotte98}.
Photoemission experiments were carried out at the Ames/Montana 
ERG/Seya beam-line at the Synchrotron Radiation Center.
The details of PES experiment are the same as those described
in Ref.~\cite{Kang01}.
The total instrumental resolution [FWHM : full width at half maximum]
was about $80$ meV and $250$ meV at $h\nu \sim 20$ eV and $h\nu 
\sim 100$ eV, respectively.
High resolution photoemission spectra were taken with the FWHM of 
about $30$ meV. 	
The photon flux was monitored by the yield from a gold mesh and 
all the spectra reported were normalized to the mesh current.
Temperature (T)-dependence of PES was also investigated below 
and above the AF transition temperature.
For T-dependent PES measurements,
the chamber pressure stayed below $\rm 7 \times 10^{-11}$ Torr 
during heating.
The low-T PES spectra were reproduced after the heating-cooling cycle. 

The electronic structures of UTSn have been calculated by employing 
the self-consistent LMTO (linearized muffin-tin-orbital) band method.
The partial densities of states (PDOS) have been calculated by using 
the LSDA+$U$ band method incorporating the spin-orbit (SO) interaction, 
so that the orbital 
polarization is properly taken into account \cite{Kwon2000}.
The von Barth-Hedin form of the exchange-correlation potential 
has been utilized. 

\begin{figure}[t]
\includegraphics[scale = 0.5]{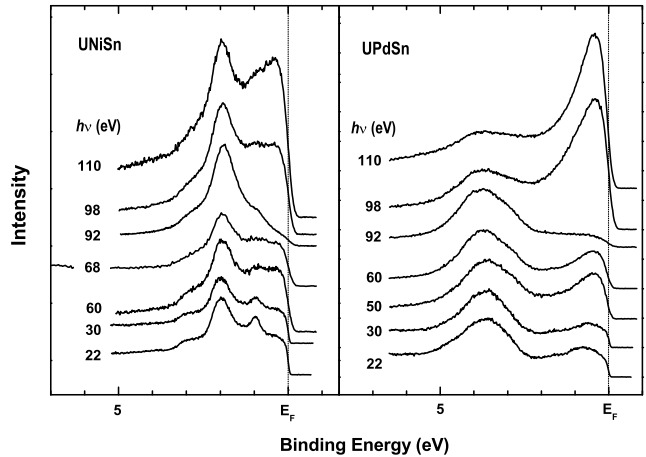}
\caption{Normalized valence-band spectra of UTSn (T=Ni, Pd).
        The $h\nu=22$ and 30 eV spectra that are arbitrarily
        scaled to show their line-shapes better.  }
\label{edc}
\end{figure}


\begin{figure}[t]
\includegraphics[scale = 0.5]{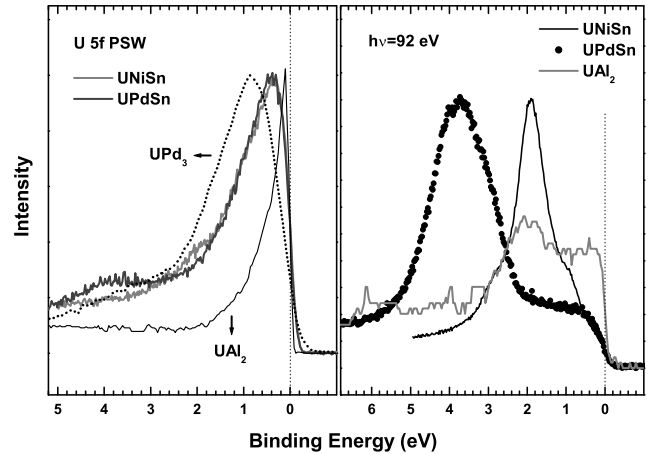}
\caption{(a) Comparison of the U $5f$ PSWs of UTSn (T=Ni, Pd),
        UAl$_2$ (Ref.\protect\cite{Allen96}),
        and UPd$_3$ (Ref.\protect\cite{Kang89}).
        (b) Comparison of the off-resonance spectra ($h\nu=92$ eV)
        of UTSn (T=Ni, Pd) and UAl$_2$ 
	(Ref.\protect\cite{Allen96})}
\label{psw}
\end{figure}

\section{Results and Discussion} 
\label{sec:results}

\subsection{U $5f$ and T $d$ PSWs} 
\label{subsec:EDC}

Figure~\ref{edc} shows the valence-band spectra of UTSn (T=Ni, Pd) 
in the photon energy ($h\nu$) range of $22-110$ eV.
At low $h\nu$'s ($22-30$ eV), 
the contribution from the Sn $sp$ electron emission is non-negligible
($10-15 \%$), and the cross-sections ($\sigma$) of U $6d$ and T $d$ 
electrons are comparable to one another ($\sim 40 \%$) \cite{Yeh85}. 
$h\nu=92$ eV, $h\nu=98$ eV, and $h\nu=110$ eV correspond to the off-
and on-resonance energies due to the U $5d_{5/2} \rightarrow 5f$
and  U $5d_{3/2} \rightarrow 5f$ absorptions, 
respectively \cite{Kang01}.
Therefore the emission enhanced at $h\nu=98$ and $h\nu=110$ eV
can be identified as the U $5f$ emission.
The off-resonance spectrum at $h\nu=92$ eV is dominated by the Ni
and Pd $d$ emission
because, at this $h\nu$, the Sn $sp$ electron
emission is negligible with respect to the T $d$ emission
($< 1 \%$ of the Ni/Pd $d$ emission) and
the U $5f$ emission is suppressed.
Using the U $5d \rightarrow 5f$ RPES, we have determined the U $5f$
PSW of UTSn. 
Before subtraction, the off-resonance spectra have been multiplied 
by a factor of 0.94 and 0.6 for UNiSn and UPdSn, respectively, 
in order to account for the $h\nu$ dependence of other conduction-band 
electrons.

Figure~\ref{psw}(a) compares the extracted $5f$ PSWs of UTSn (T=Ni, Pd)
to those of a nearly heavy fermion system UAl$_2$ \cite{Allen96}
and a typically localized $5f$ system UPd$_3$ \cite{Kang89}.
All the spectra are scaled at the peak.
UPd$_3$ is known to be a tetra-valent (U$^{4+}$) intermetallic uranium
compound with a localized $5f^2$ configuration, and so 
the $5f$ peak in UPd$_3$ is assigned as the $5f^{2} \rightarrow 5f^{1}$
transition \cite{Kang89}.
On the other hand, $5f$ electrons in UAl$_2$ is expected to be itinerant,
and so the $5f$ peak close to $\rm E_F$ in UAl$_2$ is considered to
represent the fully relaxed $5f^{n}c^{m-1}$ final states ($n=$2, 3, 4),
under the assumption of the $5f^{n}c^{m}$ ground state configuration.

Interestingly the extracted U $5f$ spectra of UNiSn and UPdSn are 
very similar each other, even though the on- and off-resonance spectra 
are very different. 
They have common features, such as a pronounced peak centered at 
about 0.3 eV binding energy (BE) and a tail to about 3 eV 
below $\rm E_F$ \cite{other}.
It is found that, as one moves from UAl$_2$ to UTSn and UPd$_3$,
the centroid of the $5f$ electron peak moves away from $\rm E_F$ 
and its width becomes wider.
This trend is accompanied by the decreasing $5f$
spectral weight at the Fermi level, N$_f$($\rm E_F$).
Compared to the large $5f$ spectral weight near $\rm E_F$ in UAl$_2$,
N$_f$($\rm E_F$) in UTSn is lower than that in UAl$_2$, suggesting
that U $5f$ electrons in UTSn are more localized then in
UAl$_2$.
This finding is consistent with the fact that UNiSn and UPdSn have 
large ordered magnetic moments ($1.55-2 \mu_B$) \cite{Kawa89,Robinson}.

The U $5f$ PSW of UNiSn is very similar to that of UPdSn.
The similarity in the U $5f$ PSW between UNiSn and UPdSn
suggests that the interaction between U $5f$ electrons in UTSn 
(T=Ni, Pd) is mediated mainly by the hybridization to conduction-band 
electrons, rather than by the direct $f$-$f$ hopping \cite{Vfd},
as explained below.
The average U-U separation $d_{U-U}$
in UPdSn ($3.63 \AA$) is much shorter than that in UNiSn 
($4.53 \AA$), but is rather close to that in UAl$_2$ ($3.22 \AA$).
$d_{U-U}=3.63 \AA$ in UPdSn lies on the border of the Hill limit
($d_{Hill}=3.3-3.5 \AA$) \cite{Hill70},
beyond which the U $5f$ electrons are observed to form local moments.
If we consider the average U-U separation only, the direct $f$-$f$ 
hopping among U $5f$ electrons is expected in UPdSn (even if it may 
be weak), while it is expected to be negligible in UNiSn.
Thus the interaction between U $5f$ electrons in UTSn should be 
mediated by the hybridization to conduction-band electrons, 
such as U $6d$, Sn $sp$, and T $d$ electrons. 
This conclusion is consistent with the fact that UNiSn and UPdSn have 
significantly larger ordered magnetic moments than in other U 
intermetallic systems \cite{Kawa89,Robinson}.
The inelastic neutron scattering study also found 
rather well-defined CEF excitations in UTSn (T=Ni, Pd) \cite{McEwen00}.

Figure~\ref{psw}(b) compares the $h\nu=92$ eV off-resonance
spectra of UNiSn and UPdSn, which can be considered to represent 
the experimental Ni $3d$ and Pd $4d$ PSWs, respectively.
To find a correlation between the U $5f$ PSW and the hybridization
effect,
we compare the off-resonance spectrum ($h\nu=92$ eV) of UAl$_2$ 
for which N$_f$($\rm E_F$) is very large.
The spectrum for UAl$_2$ is reproduced from Ref.~\cite{Allen96}
and it is scaled so that the area $\rm E_F$ and $\sim 5 $eV BE is 
comparable with that in UNiSn.
It is shown that the Pd $4d$ peak lies at a higher BE ($\sim 4$ eV BE)
than Ni $3d$ peak ($\sim 2$ eV BE) and its FWHM ($\sim 2$ eV)
is much wider than that of Ni $3d$ peak ($\sim 1$ eV). 
The latter difference reflects the less localized nature of Pd $4d$
states than Ni $3d$ states. Then the more spread wave functions
of the Pd $d$ electrons than those of the Ni $3d$ electrons would
yield the larger spatial overlap
between U $5f$ and Pd $4d$ wave functions.   
On the other hand, due to a higher BE of Pd $4d$ peak,
the energy overlap 
between U $5f$ and Pd $4d$ wave functions would be small so as to 
weaken the hybridization. It is thus expected that
the effective hybridization in UNiSn and UPdSn becomes 
more or less similar.

Note that both the Pd $4d$ and Ni $3d$ PSWs reveal a very low 
spectral intensity near $\rm E_F$, I($\rm E_F$), which is of comparable
magnitude if the main $d$ peaks are scaled at their maxima.
In contrast, UAl$_2$ reveals a much larger I($\rm E_F$) than UTSn.
This finding indicates that the reduced N$_f$($\rm E_F$) in UTSn 
arises from the energy-dependent hybridization matrix elements 
M$_{fd}(h\nu)$ between the U $5f$ states and the T $d$ states 
that have a very low DOS at $\rm E_F$.
This interpretation implies that the energy-dependent hybridization,
instead of the average hybridization strength, plays an important
role in determining N$_f$($\rm E_F$) in uranium compounds. 

\begin{figure}[t]
\includegraphics[scale = 0.65]{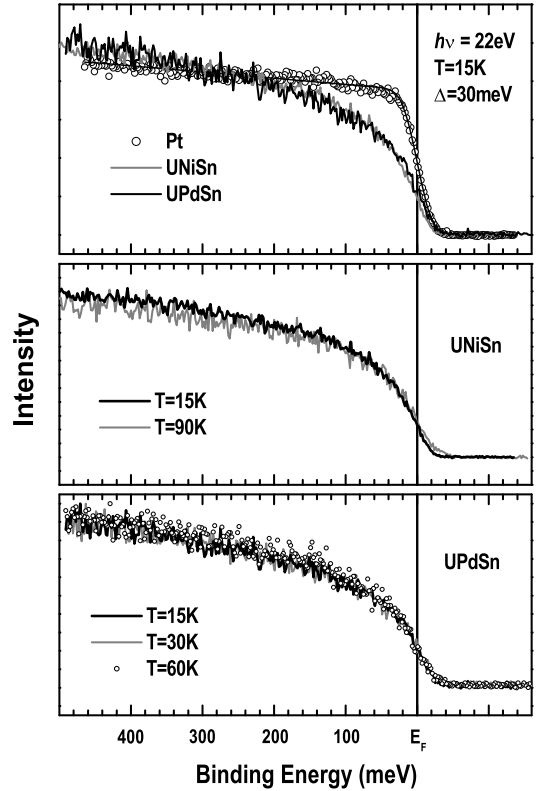}
\caption{Top: Comparison of high-resolution photoemission spectra 
	of UTSn  (T=Ni, Pd) and Pt metal in the vicinity of $\rm E_F$, 
	obtained at T$=15$ K with FWHM $\sim30$ meV.
	Middle: Comparison of the $h\nu=22$ eV spectrum of UNiSn 
	obtained at T$=15$ K (black line) to that at T$=90$ K 
	(gray line).
	Bottom: Similarly for UPdSn obtained at T$=15$ K (black line),
	T$=30$ K (gray line), and T$=60$ K (open circles).}
\label{hr}
\end{figure}
\begin{figure}[t]
\includegraphics[scale = 0.6]{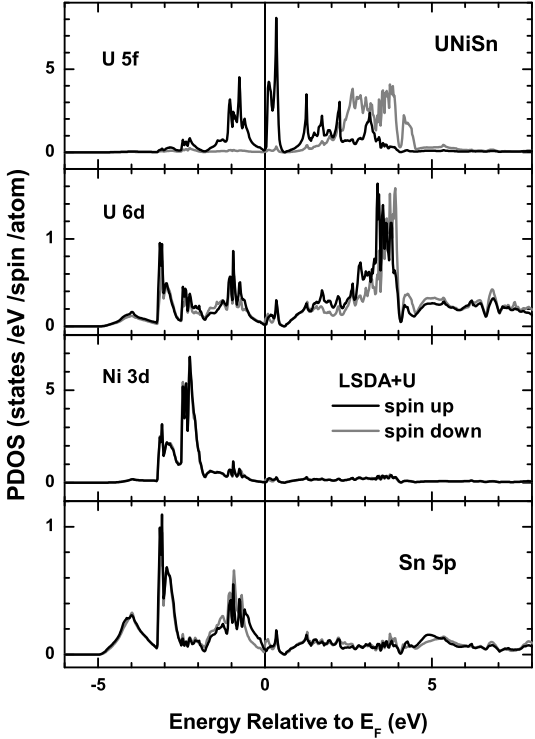}
\caption{The calculated PDOS per spin and per atom of UNiSn, 
	obtained from the LSDA$+U$ calculation for the AF ground state.
	The spin-up and spin-down PDOS are denoted with black and 
	grey lines, respectively. 
        From top are shown U $5f$, U $6d$, Ni $d$, and Sn $5p$ PDOSs. } 
\label{ulda1}
\end{figure}

\begin{figure}[t]
\includegraphics[scale = 0.58]{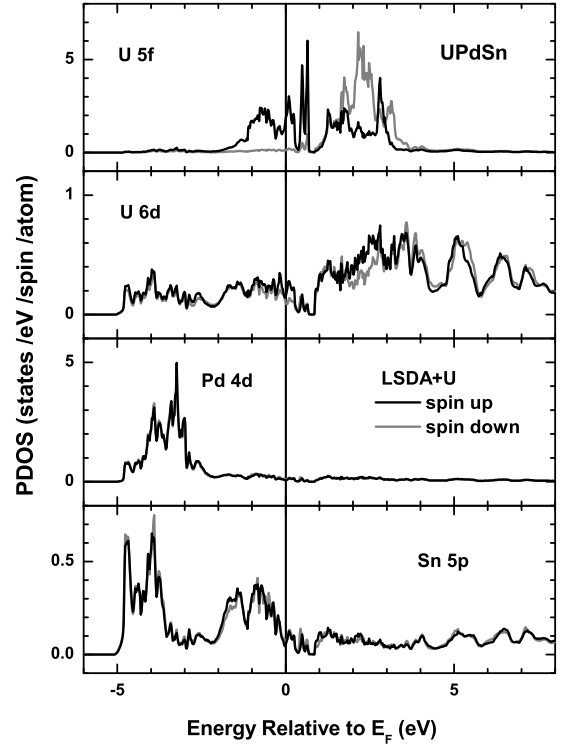}
\caption{The calculated PDOS per spin and per atom of UPdSn, 
	obtained from the LSDA$+U$ for the AF ground state. }
\label{ulda2}
\end{figure}

\subsection{High-resolution PES across phase transitions} 
\label{subsec:t-dep}

Figure~\ref{hr} shows the temperature dependence of the high-resolution PES 
spectra of UTSn (T=Ni, Pd) in the vicinity of $\rm E_F$,
obtained at $h\nu=22$ eV with FWHM $\approx 30$ meV.
All the spectra were obtained with the same measurement conditions 
except for temperature.
The top panel compares the high-resolution T=15 K PES spectra of UTSn 
and Pt metal. Pt is chosen as representing the typical metallic 
Fermi-edge spectrum. In this comparison, all the spectra are scaled 
at about 300 meV below $\rm E_F$.
The solid line along the measured spectrum of Pt metal is 
the result of the flat DOS with a non-zero slope, cut-off at $\rm E_F$ 
by the 15 K Fermi distribution function, and convoluted 
with a Gaussian function with FWHM $=30$ meV. 

Figure~\ref{hr} shows that the high-resolution PES spectra of UTSn 
in the vicinity of $\rm E_F$ are almost identical each other and that 
there is certainly a finite metallic DOS at $\rm E_F$ in UTSn.  
It is clearly shown that the slope of the PES spectrum of UTSn 
just below $\rm E_F$ is lower than that of Pt, 
indicating a lower DOS at $\rm E_F$. 
In contrast to a flat DOS with a non-zero slope for Pt, the 15 K 
spectra of UTSn are described well by the V-shaped metallic DOS 
near $\rm E_F$ \cite{Kang02}.
The V-shaped metallic DOS represents a model with a reduced but 
finite DOS at $\rm E_F$ which is usually formed in semi-metallic 
systems.
This difference confirms that UTSn have a lower DOS at $\rm E_F$ 
than a typical metal, in agreement with a low N$_f$($\rm E_F$)
(See Fig.~\ref{psw}).

The middle panel compares the spectra of UNiSn, obtained at 
$\rm T=15$K (black line), which belongs to the AF metallic phase, 
to that at $\rm T=90K$ (gray line) which belongs to the paramagnetic 
semi-conducting phase. 
Similarly, the T-dependence of the spectra of UPdSn
is shown in the bottom panel, obtained at T$=15$ K (black lines),
the monoclinic AF phase, T$=30$ K (gray lines), the orthorhombic
AF phase, and T$=60$ K (open dots), the paramagnetic phase,
respectively.
Nearly no changes have been observed in the PES spectra of UTSn 
across magnetic phase transitions, 
except that due to the temperature broadening.
Therefore the same V-shaped metallic DOS is expected to provide 
a reasonably good fit to the measured spectra for UPdSn 
at 30K, 60K, and 90 K.
Our study indicates that both UNiSn and UPdSn have finite metallic 
DOSs at $\rm E_F$ in different magnetic phases, suggesting that there 
are no appreciable changes in their electronic structures 
across the magnetic phase transitions. 

\subsection{Comparison to the LSDA+$U$ calculation} 
\label{subsec:ulda}

Figures~\ref{ulda1} and ~\ref{ulda2} show the calculated PDOSs 
per atom of UNiSn and UPdSn, respectively, obtained from the LSDA$+U$ 
calculations.
Tetragonal and orthorhombic AF structures are considered for
UNiSn and UPdSn, respectively, and the collinear spin
configurations are assumed for both systems.
The on-site Coulomb correlation parameter $U$ for the U $5f$ electrons 
is included in these calculations.
The parameters used in this calculation are the Coulomb correlation 
$U=2.0$ eV and the exchange $J=0.95$ eV and $J=0.8$ eV for UNiSn and 
UPdSn, respectively.
The on-site Coulomb correlations 
for the Ni $3d$ and Pd $4d$ electrons have been neglected.
The LSDA+$U$ yields the correct metallic ground states for the AF phase
of UTSn (T=Ni, Pd), and the correct semiconducting and metallic ground 
states for the paramagnetic phases of UNiSn and UPdSn, respectively.
The major effect of including $U$ in the LSDA+$U$ is to shift both 
the occupied $5f$ peaks and the unoccupied $5f$ peaks 
away from $\rm E_F$ \cite{Kang01}.
The second effect of the LSDA+$U$ is to shift the U $d$, T $d$,
and Sn $p$ PDOSs toward $\rm E_F$.
The larger the $U$ value is, the larger the peak shifts.
For UTSn (T=Ni, Pd), the calculated orbital and spin magnetic moments 
for U ions are $4.53 \mu_{B}$ and $-2.25 \mu_{B}$ (UNiSn),
and $4.53 \mu_{B}$ and $-2.24 \mu_{B}$ (UPdSn), respectively,
and so the total magnetic moments of U ions become 
$2.28 \mu_{B}$ (UNiSn) and  $2.29 \mu_{B}$ (UPdSn).  
These values are in reasonable agreement with experiments
\cite{Kawa89,Robinson}.
The spin magnetic moments of U $5f$ states, $\sim 2 \mu_{B}$ for both
UNiSn and UPdSn, reflect that the number of occupied $5f$-electrons
is close to two ($5f^2$) with U$^{4+}$ configuration. 
Note, however, that the $f$-electrons in UNiSn and UPdSn are not so 
localized as in UPd$_3$ due to the large hybridization with 
the neighboring elements. Therefore the $f$-electron count is not 
really meaningful in these intermetallic compounds.

\begin{figure}[t]
\includegraphics[scale = 0.57]{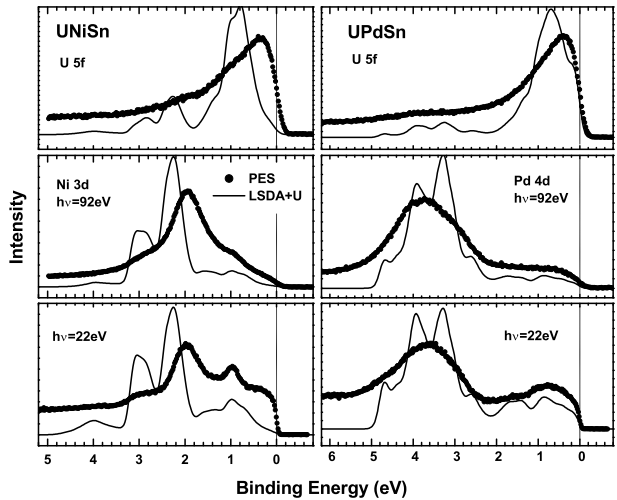}
\caption{Top: Comparison of the extracted U $5f$ PSW (dots) of UNiSn 
	(left) and UPdSn (right) to the calculated PDOS, obtained from 
	the LSDA+$U$ calculation (solid line).  
	Middle: Similarly for Ni $3d$ and Pd $4d$ states.
	Bottom: Comparison of the $h\nu=22$ eV PES spectrum of UTSn 
	(dots) to the sum of the U $6d$, T $d$, and Sn $5p$ 
	PDOSs. See the text for details.} 
\label{comp}
\end{figure}

For both UNiSn and UPdSn, the U $f$ states exhibit the exchange-split 
$5f$ bands, separated from each other by about $4$ eV and $3$ eV, 
respectively. 
The other states (U $d$, T $d$, Sn $p$) exhibit nearly no 
exchange splitting, indicating that the spin-polarization in UTSn 
is mainly due to the U $f$ electrons. 
The Ni and Pd $d$ bands are nearly filled with a very low DOS 
at $\rm E_F$, in agreement with the PES data (see Fig.~\ref{psw}).
The Sn $p$ states are spread over the whole valence band,
but relatively more concentrated at $1-2$ eV below $\rm E_F$. 
The U $d$, T $d$, and Sn $p$ PDOSs share common features, 
indicating the large hybridization among them.
The DOS at $\rm E_F$ is low for UNiSn, but high for UPdSn. 
It is because that the Fermi level in UNiSn cuts the valley of 
U $f$ DOS, while the Fermi level in UPdSn is located near the second 
peak of the U $f$ DOS. This difference arises from the different
crystal structures of UNiSn and UNiSn.

Figure~\ref{comp} compares the extracted PSWs (dots) of UTSn (T=Ni, Pd)
to the calculated PDOS, obtained from the LSDA$+U$ 
calculations (solid lines) .
In the comparison to the PES spectra, 
only the occupied part of the calculated PDOSs was taken,
and then convoluted by a Gaussian function with 0.2 eV at the FWHM. 
The Gaussian function has been used to simulate the instrumental 
resolution.  
The effects of the lifetime broadening and photoemission matrix 
elements are not included in the theory curves.
At the bottom panels, the theoretical spectra correspond to the sum
of the U $d$, T $d$, and Sn $p$ PDOS, because none of the contributions 
are negligible at $h\nu=22$ eV and 
it is difficult to separate them out
(see the discussion under Fig.~\ref{edc}).

The LSDA$+U$ calculation provides a reasonably good 
agreement with PES for UPdSn, but not so for UNiSn.
The calculated U $5f$ PDOS for UPdSn shows a metallic DOS at $\rm E_F$,
resulting in good agreement with PES.
In contrast, the calculated U $5f$ PDOS for UNiSn shows a very low DOS 
at $\rm E_F$, resulting in large disagreement with PES.
The nearly negligible DOS at $\rm E_F$ for T $d$ states (T=Ni, Pd)
in the LSDA$+U$ calculations gives good agreement with PES. 
For UPdSn, the peak positions in the LSDA$+U$ agree very well with 
the PES spectra for Pd $d$ and Sn $sp$ states. 
The most pronounced discrepancy in UNiSn is that the calculated 
peak positions in the Ni $d$ and Sn $p$ PDOSs appear at higher BEs 
than in the PES spectra by more than 0.5 eV.
The calculated  U $f$ peaks in the occupied part also appear 
at higher BEs than in PES, and the calculated N$_f$($\rm E_F$) 
is too small, as compared to PES. 
It is rather surprising 
that the LSDA$+U$ calculation for UPdSn gives
a good agreement with the measured U $5f$ PES, whereas
that for UNiSn does not.
This finding indicates that 
the more localized nature Ni $3d$ electrons, 
as compared to Pd $4d$ electrons, plays an important role 
in determining the nature of U $5f$ electrons in UNiSn, 
probably via the hybridization to U $5f$ electrons.
 
\section{Conclusions}
\label{sec:conc}

The electronic structures of UTSn (T=Ni, Pd) have been investigated by 
performing the photoemission experiment and the LSDA+$U$ 
electronic structure calculation.
The extracted U $5f$ spectra of UTSn are very similar to each other, 
showing the U $5f$ peaks at $\approx 0.3$ eV BE.
Compared to the U $5f$ PSWs of a nearly heavy fermion system UAl$_2$ 
and a typically localized $5f$ system UPd$_3$,
the centroid of the $5f$ electron peak moves away from $\rm E_F$ 
from UAl$_2$ to UTSn and UPd$_3$, accompanied by the decreasing $5f$
spectral weight at the Fermi level, N$_f$($\rm E_F$), which suggests 
that the localization of U $5f$ electrons in UTSn is 
between those in UAl$_2$ and UPd$_3$.
The similarity in the U $5f$ PSW between UNiSn and UPdSn suggests 
that the interaction between U $5f$ electrons in UTSn is mediated 
mainly by the hybridization to conduction-band electrons, rather than 
by the direct $f$-$f$ hopping.
Both the Ni $3d$ and Pd $4d$ PSWs show the main peaks well below 
$\rm E_F$ and a very low DOS at $\rm E_F$.
The high-resolution PES spectra of UTSn are also very similar
each other, with the slope just below $\rm E_F$ being lower than that 
of Pt. They are described well by a V-shaped metallic DOS 
near $\rm E_F$, consistent with the reduced $5f$ DOS at $\rm E_F$.
The temperature-dependent high-resolution PES spectra of UTSn 
do not manifest any noticeable change in their electronic structures 
across the magnetic phase transitions.  
Both the high-resolution PES and the T $d$ PSWs suggest that
the reduced N$_f$($\rm E_F$) in UTSn is ascribed to the hybridization 
to the very low T $d$ DOS at $\rm E_F$.
Our work suggests that the energy-dependent hybridization plays 
an important role in determining the U $5f$ electronic structure. 
The calculated spin magnetic moments of U $5f$ states, $\sim 2 \mu_{B}$
for both UNiSn and UPdSn, reflect that the number of occupied 
$5f$-electrons is close to two ($5f^2$) with 
the U$^{4+}$ configuration.
Comparison of the measured PES spectra to the LSDA+$U$ band structure 
calculation reveals a reasonably good agreement for UPdSn, but not for 
UNiSn, indicating the importance of the on-site Coulomb interaction 
not only for U $5f$ electrons but also for T $d$ electrons.

\acknowledgments

This work was supported by the KRF (KRF--2002-070-C00038) and
by the KOSEF through the CSCMR at SNU and the eSSC at POSTECH.
The SRC is supported by the NSF (DMR-0084402).


\begin{thebibliography}{99}
\bibitem{Vfd} The U-Ni separation ($2.77 \AA$) and the U-Pd 
	separation ($2.92 \AA$) are much smaller than 
	the U-U separations.
\end{thebibliography}
\end{document}